\documentstyle[epsf]{article}%
  \def\baselinestretch{1.8}
\begin{document}
\title{Simplified amino acid alphabets based on deviation of
        conditional probability from random background }
\author{Xin Liu${^\dagger}$, Di Liu${^\ddag}$, Ji Qi${^\dag}$,
Wei-Mou Zheng${^\dag}$\\
${^\dag}${\it Institute of Theoretical Physics, China,
Beijing 100080, China}\\
${^\ddag}${\it Center of Bioinformations at Peking University,
Beijing 100871, China}}
\date{}
\maketitle

\begin{abstract}
The primitive data for deducing the Miyazawa-Jernigan contact
energy or BLOSUM score matrix consists of pair frequency counts.
Each amino acid corresponds to a conditional probability distribution.
Based on the deviation of such conditional probability from random
background, a scheme for reduction of amino acid alphabet is proposed.
It is observed that evident discrepancy exists between reduced alphabets
obtained from raw data of the Miyazawa-Jernigan's and BLOSUM's residue
pair counts. Taking homologous sequence database SCOP40 as a test set,
we detect homology with the obtained coarse-grained substitution matrices.
It is verified that the reduced alphabets obtained well preserve
information contained in the original 20-letter alphabet.
\end{abstract}

\leftline{PACS number(s): 87.10.+e,02.50.-r}%
\bigskip

\section{Introduction}

Experimental investigation has strongly suggested that protein
folding can be achieved with fewer letters than the 20 naturally
occurring amino acids \cite{chan,plaxco}. The
native structure and physical properties of the protein Rop is
maintained when its 32-residue hydrophobic core is formed with
only Ala and Leu residues \cite{munson}. Another
example is the five-letter alphabet of Baker's group for 38 out of
40 selected sites of the SH3 chain \cite{baker}.
The mutational tolerance can be high in many regions of protein
sequences. Heterogeneity or diversity in interaction must be
present for polypeptides to have protein-like properties. However,
the physics and chemistry of polypeptide chains consisting of
fewer than 20 letters may be sufficiently simplified for a
thorough understanding of the protein folding.

A central task of protein sequence analysis is to uncover the
exact nature of the information encoded in the primary structure.
We still cannot read the language describing the final 3D fold of
an active biological macromolecule. Compared with the DNA
sequence, a protein sequence is generally much shorter, but the
size of the alphabet is five times larger. A proper coarse
graining of the 20 amino acids into fewer clusters is important
for improving the signal-to-noise ratio when extracting
information by statistical means.

Based on Miyazawa-Jernigan's (MJ) residue-residue statistical
potential \cite{mj}, Wang and Wang (WW)
reduced the alphabet \cite{ww}. They introduced a `minimal mismatch'
principle to ensure that all interactions between amino acids
belonging to any two given groups are as similar to one another as
possible. The knowledge-based MJ potential is derived from the
frequencies of contacts between different amino acid residues in a
set of known native protein structure databases. Murphy, Wallqvist
and Levy (MWL) \cite{mwl} approached the same problem using the BLOSUM
matrix derived by Henikoff and Henikoff \cite{hh}. The matrix is
deduced from amino acid pair frequencies in aligned blocks of a
protein sequence database, and is widely used for sequence
alignment and comparison.

The problem of alphabet reduction may be viewed as cluster
analysis, which is a well developed topic \cite{romes,spath}.
WW used the mismatch as an objective function
without any resemblance measure. MWL adopted a cosine-like
resemblance coefficient (with a non-standard normalization) from
the BLOSUM score matrix without any objective function, and took
the arithmetic mean of scores to define the cluster center. It is
our purpose to propose an algorithm for selecting a
reduced alphabet based on deviation of conditional probability from
random background, and to compare results obtained from different schemes
of reduction.

\section{Reduction of amino acid alphabets}

Either the MJ contact energies or BLOSUM score matrices are
deduced from the primitive frequency counts of amino acid pairs.
Taking the BLOSUM matrix as an example for specificity, following
Henikoff and Henikoff \cite{hh}, we denote the total number of amino
acid $i$, $j$ pairs $(1\le i,j \le 20)$ by $f_{ij}$. It is
convenient to introduce another set of $f'_{ij}$ with
$f'_{ij}=f_{ij}/2$ for $i\not =j$ and $f'_{ii}= f_{ii}$, which
defines a joint probability for each $i$, $j$ pair
\begin{equation}
q'_{ij}= f'_{ij}/f, \qquad f= \sum_{i=1}^{20}\sum_{j=1}^{20} f'_{ij}.
\end{equation}
The probability for the amino acid $i$ to occur is then
\begin{equation}
p_i= \sum_{j=1}^{20} q'_{ij}.
\end{equation}
Each amino acid $i$ may be described by the conditional
probability vector $\{p(j|i)\}_{j=1}^{20}$ with $p(j|i)\equiv
q'_{ij}/p_i$. This conditional probability has been used as the
attribute of amino acids in an entropic cluster analysis
\cite{zheng}.

We introduce a vector ${\bf V}^{(i)}$ to characterize amino acid
$i$. This vector has its components
\begin{equation}
V^{(i)}_j =\ln{(p(j|i)/p_j)}, \qquad j=1,2,\ldots 20,
\end{equation}
being the logarithmic odds describing the deviation of the
conditional probability $p(j|i)$ from the `random background'
$p_j$, the probability of occurrence for amino acid $j$.
($V^{(i)}_j$ is essentially the BLOSUM score.) A group $\gamma$ of
several amino acids may be described by the weighted average
vector
\begin{equation}
{\bf U}^{(\gamma)} = \frac{\sum_{i\in \gamma}p_i {\bf V}^{(i)}}
{\sum_{i\in\gamma}p_i},
\end{equation}
where the summation is taken over the amino acids in the group.

Regarding ${\bf U}^{(\gamma)}$ as the group center, the distance
of amino acid $i$ in the group from the center may be described by
$|V_j^{(i)}- U_j^{(\gamma)}|$. When we divide the 20 amino acids
into clusters, we may measure the quality of clustering with the
following weighted sum of distances
\begin{equation}
E=\sum_{\gamma} \sum_{i\in\gamma}\sum_{j=1}^{20}q'_{ij}
|V_j^{(i)}- U_j^{(\gamma)}|,
\end{equation}
which will be called the error function of clustering. For the
original 20 amino acids with each forming a cluster, we have
simply $E=0$. when amino acids are further clustered into fewer
clusters, $E$ increases. For a fixed total number $n$ of clusters,
the best clustering is obtained when value $E$ is minimized.

Starting with the amino acid pair counts of the MJ and BLOSUM
matrices, we perform simulated annealing for minimization of $E$.
The results for reduced alphabets derived from MJ and BLOSUM
counts are shown in Tables I and II, respectively.

\section{Homology detection with reduced alphabets}
It is well known that there is no generally accepted `best' method
among many existing algorithms for clustering. To evaluate the validity
of the above scheme for reduction of amino acid alphabets, we examine
whether the reduced alphabets still preserve homology of proteins.

An element of the BLOSUM matrices or BLOSUM score is defined as
$s_{ij} = \log_2 (q'_{ij}/(p_i p_j))$. Once a reduced amino acid
alphabet is found, its coarse-grained BLOSUM scores may be calculated
similarly by
\begin{equation}
s_{\gamma\delta} = \log_2 \left(\frac{\sum_{l\in\gamma}
\sum_{m\in\delta} q'_{lm}}{\left(\sum_{l\in\gamma}p_l\right)
\left(\sum_{m\in\delta} p_m\right) }\right),
\end{equation}
which is the analogue of $s_{ij}$ for clusters $\gamma$ and
$\delta$. Using such coarse-grained BLOSUM50 substitution
matrices, we perform all-against-all sequence alignment on SCOP40
database \cite{scop1,scop2} with Blast2.0 \cite{blast1,blast2}.
The gap insertion and elongation parameters used for alignment are
set to 11/1. Filter option is closed. Detection of homology, i.e.
identification of the superfamily for each sequence in the
database, is illustrated by coverage as a function of errors per
query for a set of expectation value thresholds. The coverage is
defined as the number of homologous pairs detected divided by the
total number of homologous pairs present in the database. The
error per query is defined as the total number of non-homologous
protein sequences detected with expectation value equal to or
greater than the threshold divided by the total number of aligned
sequence pairs. By varying the expectation value cutoff of
Blast2.0, the error per query value is calculated for each
clustering scheme and adjusted to 0.001 to identify homologous
sequences. The coverage as a function of the number of amino acid
clusters is shown in Fig.~1. To compare with Ref.~\cite{mwl}, in
the figure the coverage obtained with the MWL scheme is also
shown. In general, our coverage values are superior to those of
MWL.

We further study the linear regression between alignment scores
$s$ and $s'$ for homologous pairs obtained by searching with the
original and coarse-grained BLOSUM50 matrices, respectively. We
calculate the correlation coefficient $r$ and covariance $\sigma$
\begin{equation}
r=\frac{C_{ss'}}{\sqrt{C_{ss}C_{s's'}}},\qquad
\sigma = \sqrt{\frac{(1-r)^2C_{s's'}}{m-2}}
\end{equation}
with
\begin{equation}
C_{xy}=\sum_{i=1}^m (x_i-\bar{x})(y_i-\bar{y}),\qquad
\bar{x}=\hbox{$\frac 1 m$}\sum_{i=1}^m x_i,
\end{equation}
where $m$ is the sample size. The obtained correlation coefficient
and covariance as a function of number of amino acid clusters are
shown in Figs.~2 and 3, respectively. Results from the MWL scheme
are also shown for comparison.

\section{Discussion}
In the above we have proposed a scheme for amino acid alphabet
reduction based on the deviation of conditional probability from
random background. We have detected homology of sequences in SCOP
database with the derived coarse-grained BLOSUM similarity matrices.

From Tables 1 and 2, we see that the clustering using residue pair
counts of either MJ or BLOSUM is not completely hierachical. That
is, clusters formed in an early step need not be preserved in a
latter step. Such a reversal case exists for both MJ and BLOSUM,
but is rare for both.

The clustering based on MJ shows evident discrepancy from that
based on BLOSUM. For example, Tyr(Y) groups with Phe(F) in an early stage
(12 clusters) for BLOSUM, while Tyr is still separated until the stage of
2 clusters. Another example is Val(V) and Ala(A). The MJ data take
each residue in a structure into
account, whereas the BLOSUM data focus more on aligned blocks. From the
way that the pair frequency counts are obtained, the BLOSUM data are more
related to the evolutional difference of residues, while the MJ data are
related to structure difference. However, for both MJ and BLOSUM the
separation of hydrophobic and hydrophilic groups is rather clear.

It is observed that the MJ contact energies can be largely
attributed to hydrophobicity of the residue pair involved
\cite{godzik}. We see a strong correlation between our
classification based on MJ and hydrophobic values of amino acids
\cite{carl} as shown in the example
\begin{eqnarray*}
&&\left({F\atop 3.7}\ {M\atop 3.4}\ {I\atop 3.1}\ {L\atop 2.8}\
{V\atop 2.6} \right)\ \left({C\atop 2.0}\ {W\atop 1.9}\ {A\atop
1.6}\right)\ \left( {T\atop 1.2}\ {G\atop 1.0}\ {S\atop 0.6}\
{P\atop -0.2}\ {Y\atop -0.7}\ {H\atop -3.0}\ {Q\atop -4.1}\
{N\atop -4.8}\right)\\
&&\left({E\atop -8.2}\ {D\atop -9.2}\right)\ \left({K\atop -8.8}\
{R\atop -12.3}\right).
\end{eqnarray*}
Furthermore, we do see Baker's five representative letters (AIGEK)
\cite{baker} and Schafmeister's seven letters \cite{schaf}, except
for an additional cluster consisting of the extraordinary single
member Cys(C).

Our results of homology recognition with reduced alphabets
indicate that there is no significant drop in the coverage as long
as the number of letters is not smaller than 9. The percentage
coverage retained is reduced by only 10\% for 9 letters. The
correlation coefficient and covariance calculated from the linear
regression between the alignment scores obtained with the original
and coarse-grained BLOSUM matrices agree with this very well. A
strong correlation in scores is seen for number of letters not
less than 9. The correlation coefficient and covariance are still
reasonable even though the number of clusters is as small as 5. We
may conclude that the 9-letter alphabet preserves most information
of the original 20-letter alphabet, and the 5-letter alphabet is
still a reasonable choice.

\begin{quotation}
{This work was supported in part by the Special Funds for Major
National Basic Research Projects and the National Natural Science
Foundation of China.}
\end{quotation}

\def
\baselinestretch{1}

Table I. Reduced amino acid alphabets based on the
residue pair counts for MJ matrix. The first column
indicates the number of amino acid groups.

{\baselineskip=0pt
\begin{verbatim}
 2 MFILVAW      CYQHPGTSNRKDE
 3 MFILVAW      CYQHPGTSNRK        DE
 4 MFILV    ACW  YQHPGTSNRK        DE
 5 MFILV    ACW  YQHPGTSN       RK DE
 6 MFILV    A C WYQHPGTSN       RK DE
 7 MFILV    A C WYQHP   GTSN    RK DE
 8 MFILV    A C WYQHP   G TSN   RK DE
 9 MF ILV   A C WYQHP   G TSN   RK DE
10 MF ILV   A C WYQHP   G TSN   RK D E
11 MF IL  V A C WYQHP   G TSN   RK D E
12 MF IL  V A C WYQHP   G TS  N RK D E
13 MF IL  V A C WYQHP   G T S N RK D E
14 MF I L V A C WYQHP   G T S N RK D E
15 MF IL  V A C WYQ H P G T S N RK D E
16 MF I L V A C WYQ H P G T S N RK D E
\end{verbatim} }

Table II. Reduced amino acid alphabets based on the
residue pair counts for BLOSUM50 matrix. The first column
indicates the number of amino acid groups.

{\baselineskip=0pt
\begin{verbatim}
 2 IMVLFWY     GPCASTNHQEDRK
 3 IMVLFWY     GPCAST    NHQEDRK
 4 IMVLFWY     G PCAST   NHQEDRK
 5 IMVL  FWY   G PCAST   NHQEDRK
 6 IMVL  FWY   G P CAST  NHQEDRK
 7 IMVL  FWY   G P CAST  NHQED RK
 8 IMV L FWY   G P CAST  NHQED RK
 9 IMV L FWY   G P C AST NHQED RK
10 IMV L FWY   G P C A STNH    RKQE  D
11 IMV L FWY   G P C A STNH    RKQ E D
12 IMV L FWY   G P C A ST  N  HRKQ E D
13 IMV L F WY  G P C A ST  N  HRKQ E D
14 IMV L F WY  G P C A S T N  HRKQ E D
15 IMV L F WY  G P C A S T N H RKQ E D
16 IMV L F W Y G P C A S T N H RKQ E D
\end{verbatim}}

\begin{figure}
\centerline{\epsfxsize=12cm \epsfbox{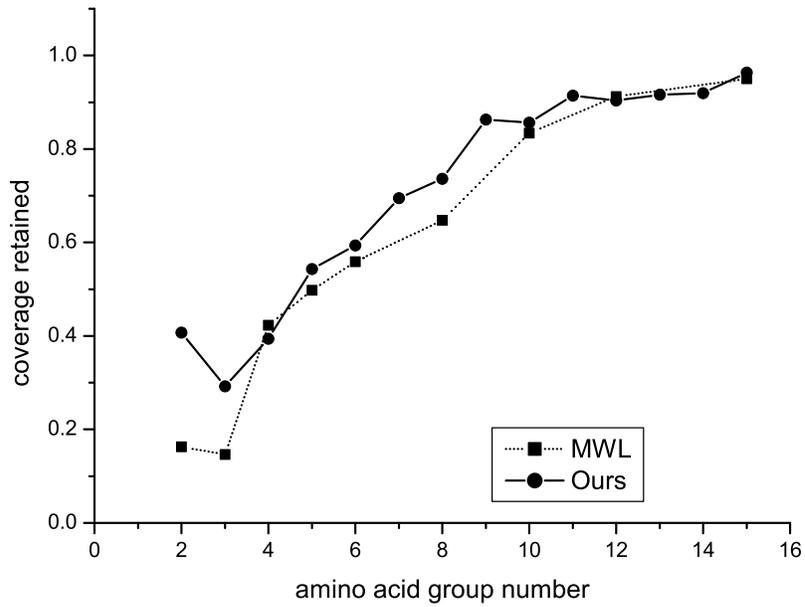}}
\caption{Retention of coverage relative to the 20-letter alphabet as
a function of the number of amino acid groups at an error
per query value of 0.001}
\label{fig1}
\end{figure}

\begin{figure}
\centerline{\epsfxsize=12cm \epsfbox{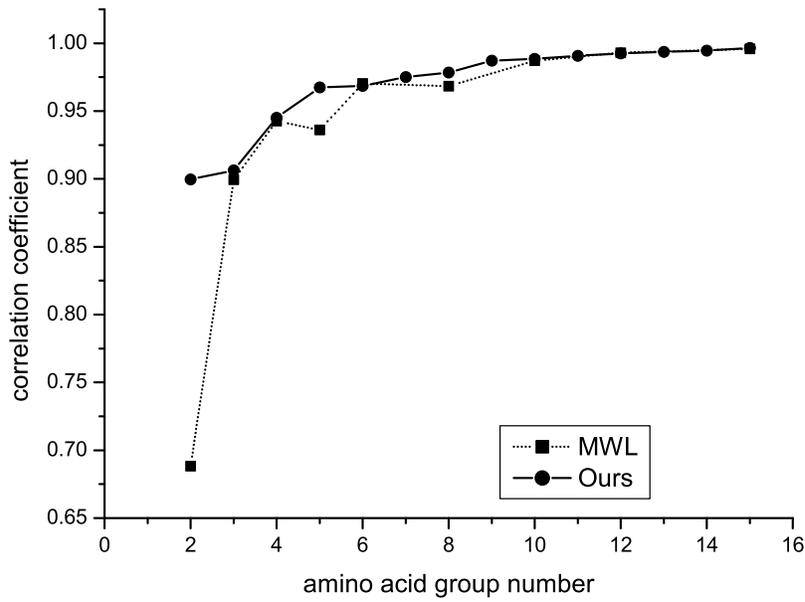}}
\caption{Correlation coefficient of linear regression between the alignment
scores obtained with the original and coarse-grained substitution matrices.
Correlation coefficient for the MWL scheme of Ref.~\cite{mwl} is also
shown for comparison.} \label{fig2}
\end{figure}

\newpage
\begin{figure}
\centerline{\epsfxsize=12cm \epsfbox{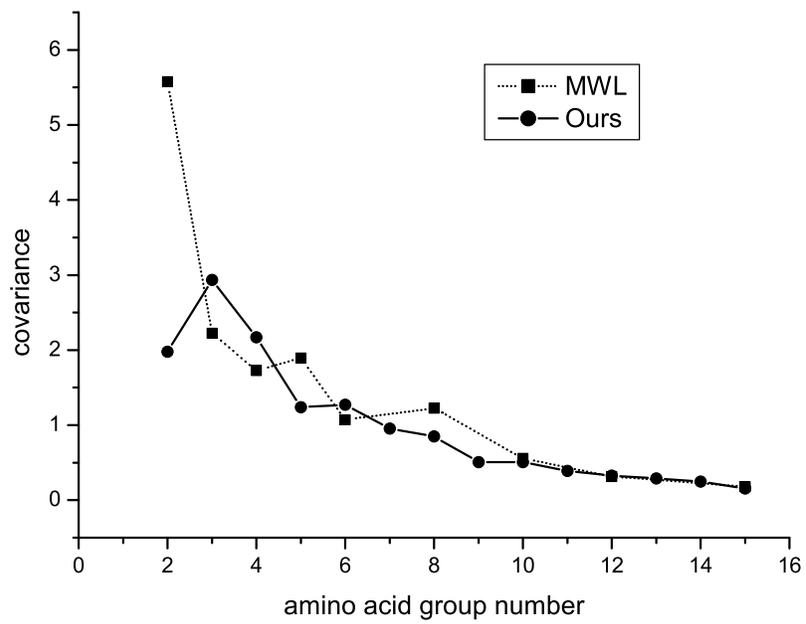}} \caption{Covariance
of linear regression between the alignment scores obtained with
the original and coarse-grained substitution matrices. Covariance
for the MWL scheme of Ref.~\cite{mwl} is also shown for
comparison. } \label{fig3}
\end{figure}

\end{document}